\documentclass[letterpaper,12pt]{article}

\pdfoutput=1
\usepackage{jheppub}
\usepackage[T1]{fontenc}
\usepackage{tikz}
\usetikzlibrary{shapes.geometric,arrows,cd,decorations.markings,intersections,hobby,arrows.meta,positioning}
\tikzset{
    baseline=(current bounding box.center),
    frozen/.style={rectangle,minimum size=0.7cm,fill=white,draw},
    mutable/.style={circle,fill=white},
    shift left/.style ={commutative diagrams/shift left={#1}},
    shift right/.style={commutative diagrams/shift right={#1}},
    decoration={
    markings,
    mark= at position 0.5 with {\node[font=\tiny] {/};}
    }
}

\DeclareMathOperator{\Gr}{Gr}
\DeclareMathOperator{\Li}{Li}

\newcommand{\AtwoOSone}{$A_2^{\text{OS}1}$}
\newcommand{\AtwoOStwo}{$A_2^{\text{OS}2}$}
\newcommand{\AtwoOS}{$A_2^{\text{OS}}$}

\catcode`@=11
\def\un#1{\relax\ifmmode\@@underline#1\else
        $\@@underline{\hbox{#1}}$\relax\fi}
\catcode`@=12

\def\m{\mu}

\def\cn{{\cal N}}

\definecolor{skyblue}{rgb}{0.12, 0.46, 1.00}
\definecolor{brightpink}{rgb}{1.0, 0.0, 0.5}
\definecolor{darkgreen}{rgb}{0.10, 0.75, 0.24}

\title{Cluster Superalgebras and Stringy Integrals}

\author[a,b]{S.~James Gates, Jr.,}
\author[a,b]{S.-N. Hazel Mak,}
\author[a,b]{Marcus Spradlin,}
\author[a]{\\Anastasia Volovich}

\affiliation[a]{Department of Physics, Brown University, Providence, RI 02912, USA}
\affiliation[b]{Brown Theoretical Physics Center, Brown University, Providence, RI 02912, USA}

\emailAdd{sylvester\_gates@brown.edu}
\emailAdd{sze\_ning\_mak@brown.edu}
\emailAdd{marcus\_spradlin@brown.edu}
\emailAdd{anastasia\_volovich@brown.edu}

\abstract{We take some initial steps to explore
physical applications of the cluster superalgebras recently
defined by Ovsienko and Shapiro.  Our primary example is a fermionic
extension of the $A_2$ cluster algebra, having fifteen cluster supervariables
instead of the usual five.
We also explore an alternate definition of cluster superalgebras
based on the promotion of cluster variables to superfields.
}

\begin{document} 
\maketitle
\flushbottom

\section{Introduction}
\label{sec:intro}

In recent years several connections have been found between
various aspects of scattering amplitudes in quantum
field theory and
cluster algebras, which were discovered by Fomin
and Zelevinsky~\cite{FZ1} in 2002 and have since been under
intense investigation by mathematicians
(see for example~\cite{clusterbk} for a  comprehensive modern
reference).

In this paper we take a few tentative steps towards
asking whether there might be any interesting
connections between \emph{super}amplitudes
and cluster \emph{super}algebras, which have very
recently begun to be explored by mathematicians~\cite{OS2015,LMRS2017,OS2018,SV2019}.

So far, the known appearances of cluster algebras
in scattering amplitudes can be organized into four
broad themes: (1) it has been observed~\cite{Golden:2013xva} that
the singularities of (certain) amplitudes are dictated
by cluster variables of the $\Gr(4,n)$ cluster
algebra;
(2) cluster structures appear naturally in
the positive Grassmannian description of
integrands~\cite{Arkani-Hamed:2012zlh} (and amplituhedra);
(3) finite-type cluster algebras provide natural
examples of ``stringy'' integrals that
generalize~\cite{Arkani-Hamed:2019mrd}
the Koba-Nielsen amplitude~\cite{Koba:1969kh}; and
(4) certain cluster polytopes are amplituhedra
for the amplitudes of bi-adjoint $\phi^3$ theory~\cite{Arkani-Hamed:2019vag}.

The first two of these four connections are
currently confined (see however~\cite{Chicherin:2020umh})
to the realm of planar
maximally supersymmetric Yang-Mills theory and
are tied, in particular, to the rich mathematical
structure of the Grassmannian $\Gr(k,n)$. It is natural
to wonder whether \emph{super} versions of
these connections could be described in terms
of cluster \emph{super}algebras associated to the
\emph{super}Grassmanian $\Gr(k|l, m|n)$ (the
space of $k|l$ planes in $\mathbb{C}^{m|n}$).
We postpone this interesting but ambitious
question to future work,
in part because the mathematics of cluster superalgebras
is not yet sufficiently well developed, though we
note that the approaches of~\cite{LMRS2017,SV2019} might
provide first steps in that direction.

Instead we largely focus on the broader and more
general third connection,
between cluster algebras and stringy integrals,
since it has interesting things to say
even for the
simplest nontrivial cluster algebra $A_2$.
We also defer consideration of the
fourth connection to future work since $\phi^3$ theory
is not amenable to supersymmetrization, although we note that
it might be interesting to look at a suitable
supersymmetrization of
$\phi^4$ theory, whose tree-level amplitudes are
geometrically encoded in the structure of
Stokes polytopes~\cite{Banerjee:2018tun}.
Finally we note that there has also been recent
interest in the connection between string
amplitudes and cluster algebras
associated to surfaces
following~\cite{Salvatori:2018fjp,Salvatori:2018aha} and
work in progress by Arkani-Hamed et al.  It would be
interesting to explore
whether there is a natural way to attach cluster
superalgebras to super Riemann surfaces (see for example~\cite{MOZ2021a,MOZ2021b}), and to connect those
to superstring amplitudes.

The structure of this paper is as follows.
In Sec.~\ref{sec:OSdef} we review the definition
of cluster superalgebras given by
Ovsienko and Shapiro in~\cite{OS2015} and study in detail
the 15 cluster supervariables associated to
the simplest nontrivial cluster superalgebra that extends the ordinary
$A_2$ algebra.
In Sec.~\ref{sec:superfield} we propose an alternate,
but perhaps more physically motivated,
definition of cluster superalgebras that is based on
promoting ordinary cluster variables to superfields.
In Sec.~\ref{sec:abel} we note that the famous five-term dilogarithm identity
remains valid when it is extended to the $A_2$ superalgebras
discussed in Secs.~\ref{sec:OSdef} and~\ref{sec:superfield}.
Finally in Sec.~\ref{sec:stringyintegrals} we explore
a few different cluster superalgebra generalizations
of the stringy integrals introduced in~\cite{Arkani-Hamed:2019mrd}.

\section{Cluster superalgebras: Ovsienko-Shapiro definition}
\label{sec:OSdef}

We assume the reader has basic familiarity with the
definition of cluster algebras in terms of
quivers and mutations \cite{FZ1,clusterbk}. 
In \cite{OS2015} Ovsienko proposed a definition of cluster superalgebras  via extended quivers and their mutations and
proved the super analog of the Laurent phenomenon.
In \cite{OS2018} Ovsienko and Shapiro refined the definition of
\cite{OS2015} and relaxed some of its constraints.
We begin this section by briefly reviewing the construction of \cite{OS2018}.

First, we define an {\it extended quiver} $\widetilde{\mathcal{Q}}$ associated to an ordinary quiver $\mathcal{Q}$, by adding:
\begin{enumerate}
\item an even number of frozen nodes with Grassmann variables $\xi_i$, and
\item one or more 2-paths which take the form $( \xi_i \rightarrow x_k \rightarrow \xi_j )$, where $x_k$ are (bosonic) variables associated to nodes of $\mathcal{Q}$.
\end{enumerate}
Note that 2-paths of opposite orientations $( \xi_i \rightarrow x_k \rightarrow \xi_j )$ and $( \xi_j \rightarrow x_k \rightarrow \xi_i )$ are not allowed. 

An example of an extended quiver is
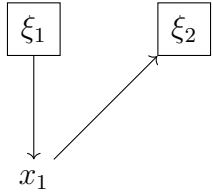
\begin{center}
\begin{tikzpicture}
\node [frozen] (f1) at (0,2) {$\xi_1$};
\node [frozen] (f2) at (2,2) {$\xi_2$};
\node (b1) at (0,0) {$x_1$};
\path[->]
(f1) edge (b1)
(b1) edge (f2);
\end{tikzpicture}
\end{center}
Here $x_1$ is a bosonic variable and $\xi_1$, $\xi_2$ are frozen Grassmann variables. Following the standard convention, boxes in the quiver diagram denote frozen nodes.
In this example there is only a single 2-path, so no ambiguity can arise,
but in general it is not enough to draw only nodes and arrows; one must
specifically indicate all 2-paths.

Next we define how to mutate an
extended quiver $\widetilde{Q}$ on node $x_k$ (this operation
will be denoted $\widetilde{\mu}_k$):
\begin{enumerate}
\setcounter{enumi}{-1}
\item The ordinary quiver $\mathcal{Q} \subset \widetilde{\mathcal{Q}}$ mutates according to the classical rules.
\item For each 2-path $( \xi_i \rightarrow x_k \rightarrow \xi_j )$: for all $( x_k \rightarrow x_l )$, add the 2-path $( \xi_i \rightarrow x_l \rightarrow \xi_j )$.
\item Reverse all 2-paths through $x_k$, i.e. change $( \xi_i \rightarrow x_k \rightarrow \xi_j )$ to $( \xi_j \rightarrow x_k \rightarrow \xi_i )$.
\item Remove any pair of 2-paths with opposite orientations, i.e.~$( \xi_i \rightarrow x_k \rightarrow \xi_j )$ and $( \xi_j \rightarrow x_k \rightarrow \xi_i )$ cancel each other.
\end{enumerate}
An example of quiver mutation is
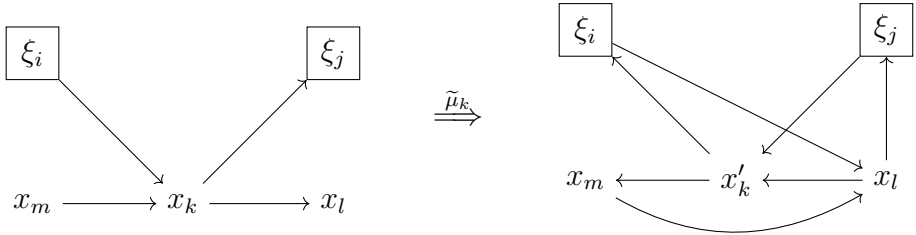
\begin{center}
\begin{tikzpicture}
\node [frozen] (f1) at (-2,2) {$\xi_i$};
\node [frozen] (f2) at (2,2) {$\xi_j$};
\node (b1) at (-2,0) {$x_m$};
\node (b2) at (0,0) {$x_k$};
\node (b3) at (2,0) {$x_l$};
\path[->]
(b1) edge (b2)
(b2) edge (b3)
(f1) edge (b2)
(b2) edge (f2);
\end{tikzpicture}
~~~~~ $\overset{\widetilde{\m}_k}{\Longrightarrow}$ ~~~~~
\begin{tikzpicture}
\node [frozen] (f1) at (-2,2) {$\xi_i$};
\node [frozen] (f2) at (2,2) {$\xi_j$};
\node (b1) at (-2,0) {$x_m$};
\node (b2) at (0,0) {$x'_k$};
\node (b3) at (2,0) {$x_l$};
\path[->]
(b2) edge (b1)
(b3) edge (b2)
(b1) edge[out=-30, in=-150] (b3)
(f2) edge (b2)
(b2) edge (f1)
(f1) edge (b3)
(b3) edge (f2);
\end{tikzpicture}
\end{center}
The mutation $\widetilde{\mu}_k$ replaces $x_k$ by $x'_k$ according to the exchange relation
\begin{equation}
    x_k x'_k ~=~ \prod_{x_k \rightarrow x_l} x_l ~+~ \prod_{\xi_i \rightarrow x_k \rightarrow \xi_j} ( 1 + \xi_i \xi_j ) \prod_{x_l \rightarrow x_k} x_l
\label{eqn:OS-exchange}
\end{equation}
leaving all other variables unchanged.

For classical cluster algebras mutation is an involution, meaning that if you mutate twice on the same node, you come back to the cluster you started with.  It is evident from the above definition that this is not the case for cluster superalgebras; indeed mutating over and over on the same node would generate an infinite number of quivers and cluster variables in general.  In order to avoid this problem
we adopt a rule of thumb whereby we
never mutate twice in a row on the same node, but instead
``walk through'' the algebra following some definite mutation
sequence adapted from the classical case.
For example, for the $A_2$ algebra we will use a mutation sequence that
alternates between the two nodes, while for the Somos-$n$ example
discussed in~\cite{clusterbk,OS2018} it was natural for them to use a cyclic mutation sequence.
It is not immediately clear how to construct
finite cluster superalgebras based on more complicated classical
algebras, such as $A_3$, where there is no suitable
choice of mutation sequence for walking through the algebra and consistently
assigning cluster supervariables to the nodes of each quiver.

In order to demonstrate these ideas we
now present two cluster superalgebras \AtwoOSone~and
\AtwoOStwo~based on $A_2$; remarkably, we will see that each contains exactly the
same 15 unique cluster supervariables.

\subsection{\texorpdfstring{Example: \AtwoOSone}{Example: A2-OS1}}
\label{subsec:OSex30}

Consider the following initial quiver with two bosonic mutable variables $x_1$, $x_2$, and two Grassmann variables $\xi_1$, $\xi_2$.
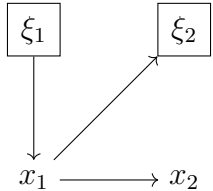
\begin{equation*}
\begin{tikzpicture}
\node [frozen] (f1) at (0,2) {$\xi_1$};
\node [frozen] (f2) at (2,2) {$\xi_2$};
\node (b1) at (0,0) {$x_1$};
\node (b2) at (2,0) {$x_2$};
\path[->]
(b1) edge (b2)
(f1) edge (b1)
(b1) edge (f2);
\end{tikzpicture}
\end{equation*}
Alternately performing the mutations $\widetilde{\m}_1$ and $\widetilde{\m}_2$
gives the sequence:
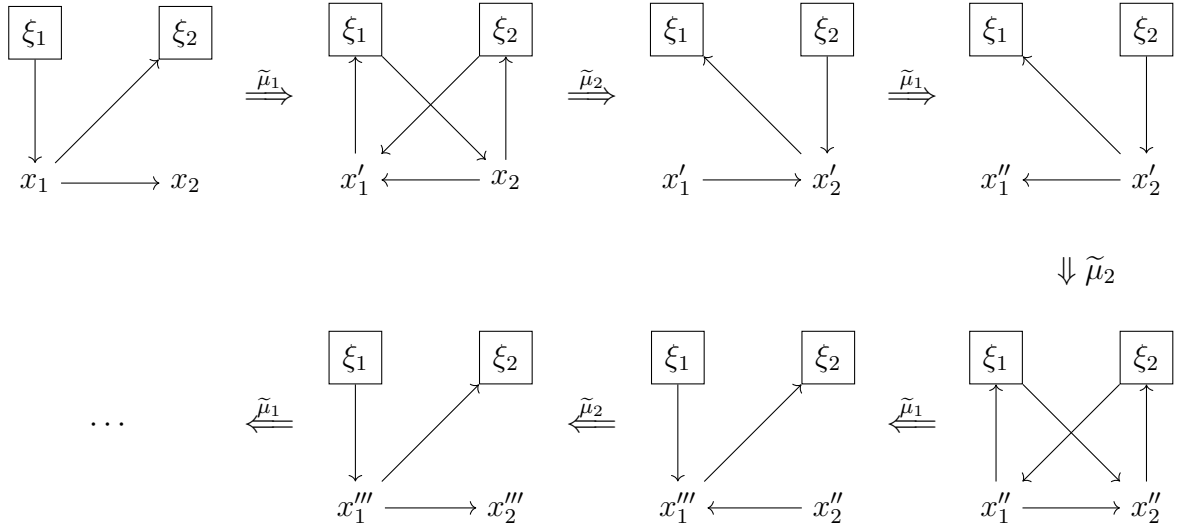
\begin{equation*}
\begin{tabular}{ccccccc}
\begin{tikzpicture}
\node [frozen] (f1) at (0,2) {$\xi_1$};
\node [frozen] (f2) at (2,2) {$\xi_2$};
\node (b1) at (0,0) {$x_1$};
\node (b2) at (2,0) {$x_2$};
\path[->]
(b1) edge (b2)
(f1) edge (b1)
(b1) edge (f2);
\end{tikzpicture} & 
$\overset{\widetilde{\m}_1}{\Longrightarrow}$ & 
\begin{tikzpicture}
\node [frozen] (f1) at (0,2) {$\xi_1$};
\node [frozen] (f2) at (2,2) {$\xi_2$};
\node (b1) at (0,0) {$x'_1$};
\node (b2) at (2,0) {$x_2$};
\path[->]
(b2) edge (b1)
(f2) edge (b1)
(b1) edge (f1)
(f1) edge (b2)
(b2) edge (f2);
\end{tikzpicture} &
$\overset{\widetilde{\m}_2}{\Longrightarrow}$ &
\begin{tikzpicture}
\node [frozen] (f1) at (0,2) {$\xi_1$};
\node [frozen] (f2) at (2,2) {$\xi_2$};
\node (b1) at (0,0) {$x'_1$};
\node (b2) at (2,0) {$x'_2$};
\path[->]
(b1) edge (b2)
(f2) edge (b2)
(b2) edge (f1);
\end{tikzpicture} &
$\overset{\widetilde{\m}_1}{\Longrightarrow}$ &
\begin{tikzpicture}
\node [frozen] (f1) at (0,2) {$\xi_1$};
\node [frozen] (f2) at (2,2) {$\xi_2$};
\node (b1) at (0,0) {$x''_1$};
\node (b2) at (2,0) {$x'_2$};
\path[->]
(b2) edge (b1)
(f2) edge (b2)
(b2) edge (f1);
\end{tikzpicture} \\[50pt]
 & & & & & & $\hspace{1em} \Downarrow \widetilde{\m}_2$ \\[15pt]
$\cdots$ & $\overset{\widetilde{\m}_1}{\Longleftarrow}$ &
\begin{tikzpicture}
\node [frozen] (f1) at (0,2) {$\xi_1$};
\node [frozen] (f2) at (2,2) {$\xi_2$};
\node (b1) at (0,0) {$x'''_1$};
\node (b2) at (2,0) {$x'''_2$};
\path[->]
(b1) edge (b2)
(f1) edge (b1)
(b1) edge (f2);
\end{tikzpicture} &
$\overset{\widetilde{\m}_2}{\Longleftarrow}$ &
\begin{tikzpicture}
\node [frozen] (f1) at (0,2) {$\xi_1$};
\node [frozen] (f2) at (2,2) {$\xi_2$};
\node (b1) at (0,0) {$x'''_1$};
\node (b2) at (2,0) {$x''_2$};
\path[->]
(b2) edge (b1)
(f1) edge (b1)
(b1) edge (f2);
\end{tikzpicture} &
$\overset{\widetilde{\m}_1}{\Longleftarrow}$ &
\begin{tikzpicture}
\node [frozen] (f1) at (0,2) {$\xi_1$};
\node [frozen] (f2) at (2,2) {$\xi_2$};
\node (b1) at (0,0) {$x''_1$};
\node (b2) at (2,0) {$x''_2$};
\path[->]
(b1) edge (b2)
(f2) edge (b1)
(b1) edge (f1)
(f1) edge (b2)
(b2) edge (f2);
\end{tikzpicture}
\end{tabular}
\end{equation*}
which, remarkably, returns to the original quiver after precisely
6 mutations.
Now let us look at the variables encountered along the way.
Using the exchange relation~(\ref{eqn:OS-exchange}), starting from $\widetilde{x}_1 \equiv x_1$ and $\widetilde{x}_2 \equiv x_2$, we have
\color{black}
\begin{equation}
\begin{split}
    \widetilde{x}_3 ~=&~ x_1' = \frac{1 + x_2}{x_1} ~+~ \frac{1}{x_1} \xi_1 \xi_2 \\
    \widetilde{x}_4 ~=&~ x_2' = \frac{1 + x_1 + x_2}{x_1 x_2} ~+~ \frac{1 + x_1}{x_1 x_2} \xi_1 \xi_2 \\
    \widetilde{x}_5 ~=&~ x_1'' = \frac{1 + x_1}{x_2} \\
    \widetilde{x}_6 ~=&~ x_2'' = x_1 ( 1 - \xi_1 \xi_2 ) \\
    \widetilde{x}_7 ~=&~ x_1''' = x_2 ( 1 - \xi_1 \xi_2 ) \\
    \widetilde{x}_8 ~=&~ x_2''' = \frac{1 + x_2}{x_1} ~+~ \frac{1}{x_1} \xi_1 \xi_2  \qquad {\rm etc.}
    \end{split}
\end{equation}

We present the resulting cluster supervariables $\widetilde{x}_n$ in Tab.~\ref{tab:OSex30}.  There we highlight the fact (manifest from~(\ref{eqn:OS-exchange})) that if we set
the odd variables to zero ($\xi_i \rightarrow 0$), they reduce to the ordinary $A_2$ cluster variables
\begin{equation}
    x_{1} ~~~,~~~
    x_{2} ~~~,~~~
    x_{3} = \frac{1 + x_{2}}{x_{1}} ~~~,~~~
    x_{4} = \frac{1 + x_{1} + x_{2}}{x_{1} x_{2}} ~~~,~~~
    x_{5} = \frac{1 + x_{1}}{x_{2}}\,.
\label{eqn:A2}
\end{equation}

\begin{table}[t]
\renewcommand{\arraystretch}{1.5}
\hspace{-1.5em}
\begin{tabular}{|c|ccccc|} \hline
    $k$ & $\widetilde{x}_{5k+1}$ & $\widetilde{x}_{5k+2}$ & $\widetilde{x}_{5k+3}$ & $\widetilde{x}_{5k+4}$ & $\widetilde{x}_{5k+5}$ \\ \hline
    0 & $x_1$ & $x_2$ & $x_3 \Big[ 1 + \frac{1}{1 + x_2} \xi_1 \xi_2 \Big]$ & $x_4 \Big[ 1 + \frac{1 + x_1}{1 + x_1 + x_2} \xi_1 \xi_2 \Big]$ & $x_5$ \\
    1 & $x_1 \Big[ 1 - \xi_1 \xi_2 \Big]$ & $x_2 \Big[ 1 - \xi_1 \xi_2 \Big]$ & $x_3 \Big[ 1 + \frac{1}{1 + x_2} \xi_1 \xi_2 \Big]$ & $x_4 \Big[ 1 + \frac{2 + 2 x_1 + x_2}{1 + x_1 + x_2} \xi_1 \xi_2 \Big]$ & $x_5 \Big[ 1 + \xi_1 \xi_2 \Big]$ \\
    2 & $x_1 \Big[ 1 - \xi_1 \xi_2 \Big]$ & $x_2 \Big[ 1 - 2 \xi_1 \xi_2 \Big]$ & $x_3 \Big[ 1 - \frac{x_2}{1 + x_2} \xi_1 \xi_2 \Big]$ & $x_4 \Big[ 1 + \frac{2 + 2 x_1 + x_2}{1 + x_1 + x_2} \xi_1 \xi_2 \Big]$ & $x_5 \Big[ 1 + 2 \xi_1 \xi_2 \Big]$ \\
    3 & $x_1$ & $x_2 \Big[ 1 - 2 \xi_1 \xi_2 \Big]$ & $x_3 \Big[ 1 - \frac{1 + 2 x_2}{1 + x_2} \xi_1 \xi_2 \Big]$ & $x_4 \Big[ 1 + \frac{1 + x_1}{1 + x_1 + x_2} \xi_1 \xi_2 \Big]$ & $x_5 \Big[ 1 + 2 \xi_1 \xi_2 \Big]$ \\
    4 & $x_1 \Big[ 1 + \xi_1 \xi_2 \Big]$ & $x_2 \Big[ 1 - \xi_1 \xi_2 \Big]$ & $x_3 \Big[ 1 - \frac{1 + 2 x_2}{1 + x_2} \xi_1 \xi_2 \Big]$ & $x_4 \Big[ 1 - \frac{x_2}{1 + x_1 + x_2} \xi_1 \xi_2 \Big]$ & $x_5 \Big[ 1 + \xi_1 \xi_2 \Big]$ \\
    5 & $x_1 \Big[ 1 + \xi_1 \xi_2 \Big]$ & $x_2$ & $x_3 \Big[ 1 - \frac{x_2}{1 + x_2} \xi_1 \xi_2 \Big]$ & $x_4 \Big[ 1 - \frac{x_2}{1 + x_1 + x_2} \xi_1 \xi_2 \Big]$ & $x_5$ \\ \hline
\end{tabular}
\caption{The 30 cluster supervariables of \AtwoOSone~one encounters along the mutation sequence described in the text.  Note that only 15 are distinct.}
\label{tab:OSex30}
\end{table}

In the table one sees some interesting patterns.
For example in each column there are only three distinct ``fermionic corrections'',
each repeated twice.  However, the rows in which they repeat
are different for different columns, but amazingly it turns out
that
\begin{equation}
    \widetilde{x}_{31} ~=~ x_1 ~~~,~~~
    \widetilde{x}_{32} ~=~ x_2 
\end{equation}
so the entire collection has a finite periodicity of
$5 \times 6 = 30$, reflecting the fact that the quiver mutation
process has a period of 6 (as mentioned above) while the bosonic
part of the variables has a period of 5.
However we note that of the 30 supervariables in the table, only 15
are distinct.

\subsection{\texorpdfstring{Example: \AtwoOStwo}{Example: A2-OS2}}
\label{subsec:OSex20}

Now consider another possible initial quiver with two even mutable variables $x_1$, $x_2$, and two odd variables $\xi_1$, $\xi_2$, with all arrows flipped with respect to previous example:
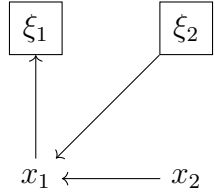
\begin{equation*}
\begin{tikzpicture}
\node [frozen] (f1) at (0,2) {$\xi_1$};
\node [frozen] (f2) at (2,2) {$\xi_2$};
\node (b1) at (0,0) {$x_1$};
\node (b2) at (2,0) {$x_2$};
\path[->]
(b2) edge (b1)
(f2) edge (b1)
(b1) edge (f1);
\end{tikzpicture}
\end{equation*}
Again we consider the alternating mutation sequence starting from $\widetilde{\m}_1$:
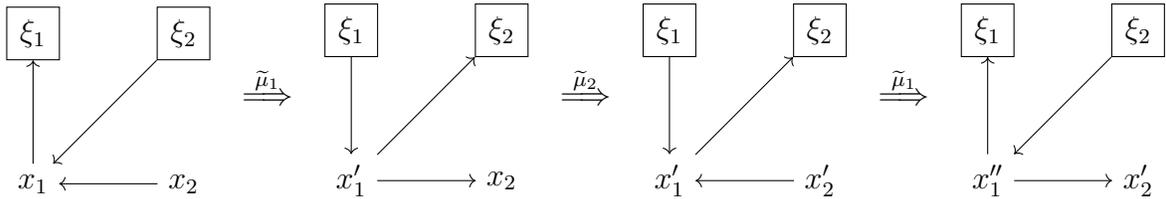
\begin{equation*}
\begin{tabular}{ccccccc}
\begin{tikzpicture}
\node [frozen] (f1) at (0,2) {$\xi_1$};
\node [frozen] (f2) at (2,2) {$\xi_2$};
\node (b1) at (0,0) {$x_1$};
\node (b2) at (2,0) {$x_2$};
\path[->]
(b2) edge (b1)
(f2) edge (b1)
(b1) edge (f1);
\end{tikzpicture} & 
$\overset{\widetilde{\m}_1}{\Longrightarrow}$ & 
\begin{tikzpicture}
\node [frozen] (f1) at (0,2) {$\xi_1$};
\node [frozen] (f2) at (2,2) {$\xi_2$};
\node (b1) at (0,0) {$x'_1$};
\node (b2) at (2,0) {$x_2$};
\path[->]
(b1) edge (b2)
(f1) edge (b1)
(b1) edge (f2);
\end{tikzpicture} &
$\overset{\widetilde{\m}_2}{\Longrightarrow}$ &
\begin{tikzpicture}
\node [frozen] (f1) at (0,2) {$\xi_1$};
\node [frozen] (f2) at (2,2) {$\xi_2$};
\node (b1) at (0,0) {$x'_1$};
\node (b2) at (2,0) {$x'_2$};
\path[->]
(b2) edge (b1)
(f1) edge (b1)
(b1) edge (f2);
\end{tikzpicture} &
$\overset{\widetilde{\m}_1}{\Longrightarrow}$ &
\begin{tikzpicture}
\node [frozen] (f1) at (0,2) {$\xi_1$};
\node [frozen] (f2) at (2,2) {$\xi_2$};
\node (b1) at (0,0) {$x''_1$};
\node (b2) at (2,0) {$x'_2$};
\path[->]
(b1) edge (b2)
(f2) edge (b1)
(b1) edge (f1);
\end{tikzpicture}
\end{tabular}
\end{equation*}
and applying $\widetilde{\mu}_2$ to the fourth quiver brings us
back to the first.
The quiver period is therefore 4, while the bosonic $A_2$ period is still 5; therefore the overall period for this cluster superalgebra is $4 \times 5 = 20$.
The cluster supervariables are listed in Tab.~\ref{tab:OSex20},
and it is clear that taking the $\xi_i \rightarrow 0$ limit
reduces to 4 copies of the classical $A_2$ algebra.
There is one repeated variable
in each column, so the total number of distinct variables
is 15. In fact these are the precisely the same as the 15 variables
of \AtwoOSone!

\begin{table}[t]
\renewcommand{\arraystretch}{1.5}
\hspace{-1.5em}
\begin{tabular}{|c|ccccc|} \hline
    $k$ & $\widetilde{f}_{5k+1}$ & $\widetilde{f}_{5k+2}$ & $\widetilde{f}_{5k+3}$ & $\widetilde{f}_{5k+4}$ & $\widetilde{f}_{5k+5}$ \\ \hline
    0 & $x_1$ & $x_2$ & $x_3 \Big[ 1 - \frac{x_2}{1 + x_2} \xi_1 \xi_2 \Big]$ & $x_4 \Big[ 1 - \frac{x_2}{1 + x_1 + x_2} \xi_1 \xi_2 \Big]$ & $x_5 \Big[ 1 + \xi_1 \xi_2 \Big]$ \\
    1 & $x_1 \Big[ 1 + \xi_1 \xi_2 \Big]$ & $x_2 \Big[ 1 - \xi_1 \xi_2 \Big]$ & $x_3 \Big[ 1 - \frac{1 + 2 x_2}{1 + x_2} \xi_1 \xi_2 \Big]$ & $x_4 \Big[ 1 + \frac{1 + x_1}{1 + x_1 + x_2} \xi_1 \xi_2 \Big]$ & $x_5 \Big[ 1 + 2 \xi_1 \xi_2 \Big]$ \\
    2 & $x_1$ & $x_2 \Big[ 1 - 2 \xi_1 \xi_2 \Big]$ & $x_3 \Big[ 1 - \frac{x_2}{1 + x_2} \xi_1 \xi_2 \Big]$ & $x_4 \Big[ 1 + \frac{2 + 2 x_1 + x_2}{1 + x_1 + x_2} \xi_1 \xi_2 \Big]$ & $x_5 \Big[ 1 + \xi_1 \xi_2 \Big]$ \\
    3 & $x_1 \Big[ 1 - \xi_1 \xi_2 \Big]$ & $x_2 \Big[ 1 - \xi_1 \xi_2 \Big]$ & $x_3 \Big[ 1 + \frac{1}{1 + x_2} \xi_1 \xi_2 \Big]$ & $x_4 \Big[ 1 + \frac{1 + x_1}{1 + x_1 + x_2} \xi_1 \xi_2 \Big]$ & $x_5$ \\ \hline
\end{tabular}
\caption{The 20 cluster supervariables of \AtwoOStwo; the
15 unique entries in this table are precisely the same as
the 15 unique entries in Tab.~\ref{tab:OSex30}.}
\label{tab:OSex20}
\end{table}

\section{Cluster superalgebras: superfield definition}
\label{sec:superfield}

In this section, we will propose a different construction of cluster superalgebras by promoting cluster variables to superfields.
We consider $\mathcal{N}=1$ superfunctions
\begin{equation}
    X_{k} ~=~ a_k ( 1 ~+~ \eta \, \theta_k )
\end{equation}
where $\eta$ is a fixed Grassmann parameter and $\theta_k$
is the Grassmann partner of $a_k$; the fact that the overall
factor of $a_k$ is pulled out is a convenient choice of normalization
(we adopted the same convention already in Tables~\ref{tab:OSex30}
and~\ref{tab:OSex20}).

We propose to apply the ordinary cluster algebra exchange
relation to superfields:
\begin{equation}
    X_k X'_k ~=~ \prod_{i \rightarrow k} X_i ~+~ \prod_{k \rightarrow i} X_i \,.
    \label{eq:supermutation}
\end{equation}
This can be expanded into individual exchange relations
\begin{align}
    a_k a'_k ~=&~ \prod_{i \rightarrow k} a_i + \prod_{k \rightarrow i} a_i \\
    a_k a'_k \, ( \theta_k + \theta'_k ) ~=&~ \left( \prod_{i \rightarrow k} a_i \right) \left( \sum_{i \rightarrow k} \theta_i \right) + \left( \prod_{k \rightarrow i} a_i \right) \left( \sum_{k \rightarrow i} \theta_i \right)
\end{align}
which define $a_k'$ and $\theta_k'$ in terms of the unprimed variables.
The exchange relation for the bosonic components $a_k$ is
identical to that of classical cluster algebras.

We could also consider $\cn=2$ superfunctions
of the form
\begin{equation}
    X_k ~=~ a_k ( 1 ~+~ \eta_1 \, \theta_k ~+~ \eta_2 \, \tilde{\theta}_k ~+~ \eta_1 \eta_2 \, z_{k} )\,.
\end{equation}
In this case the exchange relation~(\ref{eq:supermutation})
can be expanded into four component exchange relations.
The relations for $a_k$, $\theta_k$ and $\tilde{\theta}_k$
would be the same as in the $\mathcal{N}=1$ case, while
the mutation of $z_k$ would be governed by
\begin{align}
\begin{split}
    a_k a'_k \, ( z_k + z'_k - \theta_k \tilde{\theta}'_k - \theta'_k \tilde{\theta}_k ) ~=&~ \left( \prod_{i \rightarrow k} a_i \right) \sum_{i \rightarrow k} \left( z_i - \sum_{j (\rightarrow k) \neq i} \theta_i \tilde{\theta}_j \right) \\
    &~+ \left( \prod_{k \rightarrow i} a_i \right) \sum_{k \rightarrow i} \left( z_i - \sum_{(k \rightarrow) j \neq i} \theta_i \tilde{\theta}_j \right).
\end{split}
\end{align}
Of course one could just as easily also consider higher $\mathcal{N}$.
Unlike the OS definition of cluster superalgebras
reviewed in Sec.~\ref{sec:OSdef}, it is manifest that
applying the classical exchange relation to superfields makes
mutation an involution.  The superfield construction therefore
provides a family (indexed by $\mathcal{N}$) of manifestly
finite cluster superalgebras to any finite classical cluster algebra.
We now work out two examples; some others are given in the appendix.

\subsection{\texorpdfstring{Example: $A_1^{\text{SF}}$}{Example: A1-SF}}
\label{subsec:SFexA1}

The $A_1$ cluster variables are
\begin{equation}
    X_1 ~~~~~~ X_2 = \frac{1}{X_1}
\end{equation}
in the classical case.  Promoting these to
$\mathcal{N}=1$ superfields gives the cluster
supervariables
\begin{equation}
\begin{split}
    X_{1} ~=&~ a_1 ( 1 ~+~ \eta \, \theta_1 ) \\
    X_{2} ~=&~ \frac{1}{a_1} (1 ~-~ \eta \, \theta_1 )\,,
\end{split}
\end{equation}
while promoting them to $\mathcal{N}=2$ superfields gives
\begin{equation}
\begin{split}
    X_{1} ~=&~ a_1 ( 1 ~+~ \eta_1 \, \theta_1 ~+~ \eta_2 \, \tilde{\theta}_1 ~+~ \eta_1 \eta_2 \, z_1 ) \\
    X_{2} ~=&~ \frac{1}{a_1} \, \Big(~ 1 ~-~ \eta_1 \, \theta_1 ~-~ \eta_2 \, \tilde{\theta}_1 ~-~ \eta_1 \eta_2 \, ( z_1 ~+~ 2 \, \theta_1 \tilde{\theta}_1 ) ~\Big)\,.
\end{split}
\end{equation}

\subsection{\texorpdfstring{Example: $A_2^{\text{SF}}$}{Example: A2-SF}}
\label{subsec:SFexA2}

The classical $A_2$ cluster variables are
\begin{equation}
    X_{1} ~~~,~~~
    X_{2} ~~~,~~~
    X_{3} = \frac{1 + X_{2}}{X_{1}} ~~~,~~~
    X_{4} = \frac{1 + X_{1} + X_{2}}{X_{1} X_{2}} ~~~,~~~
    X_{5} = \frac{1 + X_{1}}{X_{2}} \,.
\end{equation}
If we write cluster variables in the form $X_i = a_i ( 1 + \eta \, \theta_i)$, the bosonic components $a_i$ would be the $A_2$ cluster variables, and the fermionic components are
\begin{equation}
\begin{split}
    \theta_3 ~=&~ -~ \theta_1 ~+~ \frac{a_2}{1 + a_2} \, \theta_2 \\
    \theta_4 ~=&~ -~ \frac{1 + a_2}{1 + a_1 + a_2} \, \theta_1 ~-~ \frac{1 + a_1}{1 + a_1 + a_2} \, \theta_2 \\
    \theta_5 ~=&~ \frac{a_1}{1 + a_1} \, \theta_1 ~-~ \theta_2\,.
\end{split}
\label{eq:thetas}
\end{equation}

For the $\mathcal{N}=2$ case, if we let $X_i = a_i ( 1 + \eta_1 \, \theta_i + \eta_2 \, \tilde{\theta}_i + \eta_1 \eta_2 \, z_i )$,
we find the same $a_k$'s as in the classical case,
two copies of~(\ref{eq:thetas}) (one copy for $\theta$ and
one for $\tilde{\theta})$, and finally
\begin{equation}
\begin{split}
    z_3 ~=&~ -~ z_1 ~+~ \frac{a_2}{1 + a_2} \, z_2 ~-~ 2 \, \theta_1 \tilde{\theta}_1 ~+~ \frac{a_2}{1 + a_2} \, \big( \theta_1 \tilde{\theta}_2 + \theta_2 \tilde{\theta}_1 \big) \\
    z_4 ~=&~ -~ \frac{1 + a_2}{1 + a_1 + a_2} \, z_1 ~-~ \frac{1 + a_1}{1 + a_1 + a_2} \, z_2 ~-~ \frac{2 (1 + a_2)}{1 + a_1 + a_2} \, \theta_1 \tilde{\theta}_1 ~-~ \frac{2 (1 + a_1)}{1 + a_1 + a_2} \, \theta_2 \tilde{\theta}_2 \\
    &~-~ \frac{1}{1 + a_1 + a_2} \, \big( \theta_1 \tilde{\theta}_2 + \theta_2 \tilde{\theta}_1 \big) \\
    z_5 ~=&~ \frac{a_1}{1 + a_1} \, z_1 ~-~ z_2 ~-~ 2 \, \theta_2 \tilde{\theta}_2 ~+~ \frac{a_1}{1 + a_1} \, \big( \theta_1 \tilde{\theta}_2 + \theta_2 \tilde{\theta}_1 \big)\,.
\end{split}
\end{equation}

\section{A super cluster polylogarithm identity}
\label{sec:abel}

Here we pause to consider one
aspect of the connection between cluster algebras and amplitudes
that emerged from the study of planar $\mathcal{N}=4$
super-Yang-Mills (pSYM) theory.
Namely, cluster algebras have provided an important tool for identifying
and elucidating the many nontrivial functional relations
satisfied by multiple polylogarithm functions.
The simplest of these is the remarkable relation
\begin{align}
\label{eq:abelidentity}
\sum_{i=1}^5 \left[\Li_2(-x_i) + \ln x_i \, \ln x_{i+1} + \frac{\pi^2}{10}\right] = 0\
\end{align}
for the dilogarithm function
\begin{align}
\Li_2(z) = - \int_0^z \frac{dt}{t} \ln(1-t)\,.
\end{align} 
In~(\ref{eq:abelidentity}) the sum is taken over the five cluster variables $x_i$
of the $A_2$ cluster algebra defined by the exchange relation
\begin{align}
\label{eq:A2exchange}
1 + x_i = x_{i-1} x_{i+1}
\end{align}
(which implies that $x_{i+5}=x_i$).
The identity~(\ref{eq:abelidentity}) is equivalent to a form
attributed to Abel~\cite{Abel}, though the geometric properties of the
pentagram of
arguments were studied already by Gauss~\cite{Gauss}.
In fact~(\ref{eq:abelidentity}) is the \emph{only} non-trivial
identity for the dilogarithm, in the
sense that every other identity one can write
is a functional consequence of it and the ``trivial'' identities that
relate $\Li_2(1-x)$ or $\Li_2(-1/x)$ to $\Li_2(x)$.

Polylogarithm identities involving cluster variables
emerge naturally from the study of
perturbative scattering amplitudes in pSYM theory.  For example, by expressing its 2-loop 7-particle MHV amplitude in two different ways, guaranteed to be equal to each other as a simple
physical consequence of
parity symmetry, the authors of~\cite{Golden:2013xva}
discovered a mathematically
nontrivial 40-term functional equation for the trilogarithm
function $\Li_3(z) = \int_0^z \frac{dt}{t} \Li_2(t)$
whose arguments are cluster Poisson coordinates on the moduli
space of 6 cyclically ordered points in $\mathbb{P}^2$ (the
$D_4$ cluster algebra).
More generally, numerous
identities at various weights
have emerged from the study of Feynman integrals
in quantum field theory; for recent developments
see for example~\cite{Gangl,CGR}
and references therein.
In a recent mathematical breakthrough, Goncharov and Rudenko
have used the link between cluster varieties
and polylogarithms to prove Zagier's polylogarithm conjecture
in weight 4~\cite{GR}.

It is therefore natural to ask whether the cluster superalgebras
we have explored in the previous sections have any interesting
implications for polylogarithm identities.
Interestingly, it is easy to check that the
Abel identity~(\ref{eq:abelidentity})
remains valid if we take the sum over 30 terms, with the $x_i$
consisting of the variables from Sec.~\ref{subsec:OSex30};
or over 20 terms using the $x_i$ from Sec.~\ref{subsec:OSex20}; or over the
15 unique cluster supervariables from either of those two lists.

A moment's reflection reveals that in
fact~(\ref{eq:abelidentity}) remains satisfied under
\emph{any} first-order infinitesimal deformation of the cluster
variables.
If we take $x_i \to x_i + \epsilon$ (for any single individual $x_i$),
then~(\ref{eq:abelidentity}) changes by the amount
\begin{align}
- \frac{\epsilon}{x_i} \ln(1 + x_i) + \frac{\epsilon}{x_i} \ln x_{i+1} + \frac{\epsilon}{x_{i}} \ln x_{i-1} + {\cal O}(\epsilon^2) = 0 + {\cal O}(\epsilon^2)
\label{eq:expansion}
\end{align}
by virtue of~(\ref{eq:A2exchange}).  This simple
calculation suggests that the Abel identity remains
satisfied by \emph{any} fermionic extension
of the $A_2$ cluster algebra, for which the higher
terms in~(\ref{eq:expansion}) would automatically vanish
by Grassmann statistics.

However, in general the second (and higher) order terms
in~\ref{eq:expansion} would not vanish unless
the $x_i$ are deformed in a way that conspires to produce
miraculous cancellation between various terms in the sum.
One way to guarantee this nontrivial
cancellation is to take the $x_i$
to be superfields (with arbitrary $\mathcal{N}$) satisfying~(\ref{eq:A2exchange}), as suggested
in Sec.~\ref{sec:superfield}.  It would be interesting
to investigate whether there are any other
``$\mathcal{N}>1$''   fermionic extensions of $A_2$ that preserve
the Abel identity.  Finally, of course it would be very interesting
to study fermionic extensions of
higher cluster polylogarithm identities such as the 40-term
$\Li_3$ identity from~\cite{Golden:2013xva}.

\section{Super-stringy integrals}
\label{sec:stringyintegrals}

In this section we turn our attention to the stringy
integrals defined in~\cite{Arkani-Hamed:2019mrd}, specifically
those associated to finite cluster algebras.
For an algebra ${\cal A}$ of rank $d$, the associated
stringy integral is defined by
\begin{align}
{\cal I}_{\cal A}=
(\alpha')^d \int_0^\infty \prod_{i=1}^d \frac{d y_i}{y_i}
y_i^{\alpha' X_i} \prod_{j} (F_j({\bf{y}}))^{-\alpha' c_j}
\label{eq:stringyintegral}
\end{align}
where ${\bf y}=(y_1,\ldots,y_d)$,
$\alpha'$ and the $c_j$'s are positive real parameters
and the $F_j$ are the $F$-polynomials~\cite{FZ4} of the algebra (to
be reviewed shortly).

The structure of the integral ${\cal I}_{\cal A}$
is naturally encoded in a polytope ${\cal P}_{\cal A}$
defined as follows.
If we let ${\bf N}[F]$ denote
the Newton polytope in $\mathbb{R}^d$ associated to a polynomial
$F({\bf y})$, then ${\cal P}_{\cal A}$ is the Minkowksi sum
(over $j$) of $c_j {\bf N}[F_j]$.
The main results of~\cite{Arkani-Hamed:2019mrd} are
\begin{enumerate}
\item The integral ${\cal I}_{\cal A}$ converges if and only
if $(X_1,\ldots,X_d)$ lies inside ${\cal P}_{\cal A}$, and
\item $\lim_{\alpha' \to 0} {\cal I}_{\cal A}$ is the canonical
function associated to ${\cal P}_{\cal A}$.
\end{enumerate}
The convergence criterion requires ${\cal P}_{\cal A}$
to be full-dimensional, and the canonical function is the coefficient
of the $\mathbb{R}^d$ top-form in the canonical
form~\cite{Arkani-Hamed:2019plo} associated to ${\cal P}_{\cal A}$.

Integrals of the type~(\ref{eq:stringyintegral}) are of interest
to physicists because they generalize the classic tree-level
Koba-Nielsen string scattering formula~\cite{Koba:1969kh} in a
way that manifests factorization at arbitrary $\alpha'$.

Now let us review the definition of $F$-polynomials, since this
will enable a generalization of~(\ref{eq:stringyintegral})
to cluster superalgebras.
Given an initial quiver for the cluster algebra ${\cal A}$, we
add, for every mutable node $x_i$, a frozen node labeled by
a coefficient $y_i$ and an arrow $y_i \to x_i$.
The $F$-polynomials are then the cluster variables with all $x_i$ set
to 1.  (We don't include the trivial $F$-polynomials associated
to the cluster variables in the initial quiver, since these are just $1$.)

$F$-polynomials for cluster superalgebras can be defined
in the same way and will, for the cases we consider (with
only two fermionic variables $\xi_1, \xi_2$), always take the form
\begin{equation}
    F_j({\bf y}) ~~~~~\longrightarrow~~~~~ F_j({\bf y}, \xi_1,\xi_2) \equiv F_j({\bf y}) ~ \Big[ 1 + G_j({\bf y})\, \xi_1 \xi_2 \Big]
\end{equation}
where the right-hand side defines the quantities $G_j({\bf y})$.
For a cluster superalgebra with more than two
fermions, there would in general be several additional
possible structures inside the brackets.

Compared to the standard (bosonic) stringy integral,
when we pass to the superalgebra
the term in the stringy integrand involving $F_j({\bf y})$
therefore picks up the factor
\begin{equation}
    \Big[ 1 + G_j({\bf y})\, \xi_1 \xi_2\Big]^{-\alpha' c} = 1 - \alpha' c_j ~ G_j({\bf y})\, \xi_1 \xi_2\,.
\end{equation}
Since the terms involving fermions are all manifestly proportional
to $\alpha'$, $\lim_{\alpha' \to 0} {\cal I}_{\cal A}$ would be
completely unchanged, compared to the bosonic case, if we made
no other modifications.

Instead we propose to study a slight modification of~(\ref{eq:stringyintegral}) that probes the ${\cal O}(\alpha')$ effects
of the fermionic contributions.  Our working definition of the
super-stringy integral for an OS-type cluster superalgebra
${\cal A}$ with two fermions $\xi_1,\xi_2$
is
\begin{equation}
    \mathcal{I}_{\cal{A}}^f = (\alpha')^{d-1} \int d\xi_1d\xi_2~ \int_{0}^\infty \prod_{i=1}^d \frac{dy_i}{y_i}  y_i^{\alpha'  X_i} \prod_{j}(F_j({\bf y}, \xi_1, \xi_2))^{-\alpha' c_j}\,.
\label{eqn:string-modified}
\end{equation}
For the superfield-type cluster superalgebras
we will see in Sec.~\ref{sec:A2SF} that instead of a 2-fold integral over
both $\xi_1$ and $\xi_2$, it is more natural to
look  at 1-fold integrals over each of the $\xi_i$ separately.

\subsection{\texorpdfstring{$A_2$}{A2}}
\label{subsec:stringy-A2}

For the purpose of review let us consider
first the classic $A_2$ example before moving to
cluster superalgebras.
The $F$-polynomials are
\begin{equation}
    F_3 = 1 + y_1 ~~~,~~~
    F_4 = 1 + y_1 + y_1 y_2 ~~~,~~~
    F_5 = 1 + y_2
\label{eqn:A2Fpoly}
\end{equation}
and the stringy integral is therefore
\begin{equation}
\begin{split}
    \mathcal{I}_{A_2}
    =
     (\alpha')^2 \int_0^\infty \frac{d y_1}{y_1} \, \frac{d y_2}{y_2} ~ y_1^{\alpha' X_1 } y_2^{\alpha' X_2 } 
     (1 + y_1)^{-\alpha' c_3} (1 + y_1 + y_1 y_2)^{-\alpha' c_4} (1 + y_2)^{-\alpha' c_5}\,.
\end{split}
\label{eq:pentagonintegral}
\end{equation}
The Newton polytopes associated to $c_3 F_3$, $c_4 F_4$ and
$c_5 F_5$ are respectively
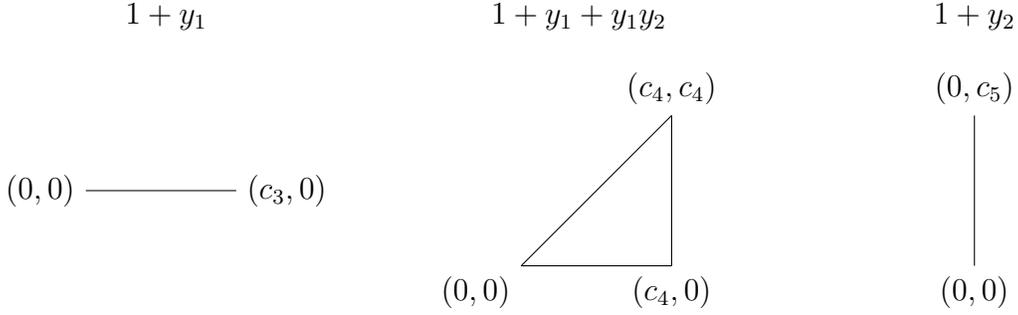
\begin{center}
\begin{tabular}{ccc}
$1 + y_1$ ~~~~~ & $1 + y_1 + y_1 y_2$ ~~~~~~~~~~~~~~~ & $1 + y_2$ \\[10pt]
\begin{tikzpicture}
\coordinate [label={left:$(0,0)$}] (p1) at (0,0);
\coordinate [label={right:$(c_3, 0)$}] (p2) at (2,0);
\path[-]
(p1) edge (p2);
\end{tikzpicture}
~~~~~ & 
\begin{tikzpicture}
\coordinate [label={below left:$(0,0)$}] (p1) at (0,0);
\coordinate [label={below:$(c_4, 0)$}] (p2) at (2,0);
\coordinate [label={above:$(c_4, c_4)$}] (p5) at (2,2);
\path[-]
(p1) edge (p2)
(p2) edge (p5)
(p1) edge (p5);
\end{tikzpicture}
~~~~~~~~~~~~~~~ &
\begin{tikzpicture}
\coordinate [label={below:$(0,0)$}] (p1) at (0,0);
\coordinate [label={above:$(0,c_5)$}] (p3) at (0,2);
\path[-]
(p1) edge (p3);
\end{tikzpicture}
\end{tabular}
\end{center}
and their Minkowski sum is the pentagon
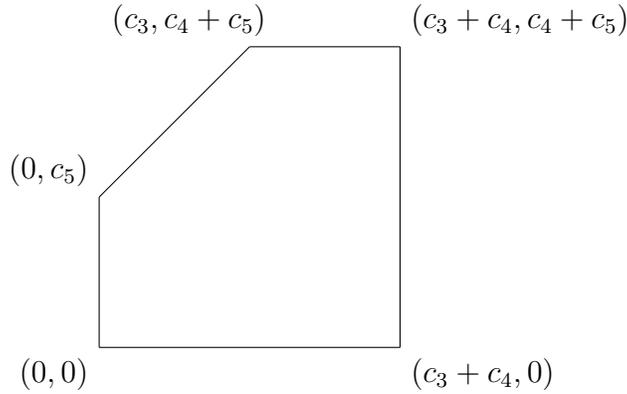
\begin{center}
\begin{tikzpicture}
\coordinate [label={below left:$(0,0)$}] (p1) at (0,0);
\coordinate [label={below right:$(c_3 + c_4, 0)$}] (p2) at (4,0);
\coordinate [label={above left:$(0,c_5)$}] (p3) at (0,2);
\coordinate [label={[xshift=2ex]above left:$(c_3, c_4 + c_5)$}] (p4) at (2,4);
\coordinate [label={above right:$(c_3 + c_4, c_4 + c_5)$}] (p5) at (4,4);
\path[-]
(p1) edge (p2)
(p1) edge (p3)
(p2) edge (p5)
(p3) edge (p4)
(p4) edge (p5);
\end{tikzpicture}
\end{center}
The integral converges for $(X_1,X_2)$ taking values in the interior
of this pentagon, and
the $\alpha'\to 0$ limit gives the canonical function associated to this pentagon:
\begin{multline}
\lim_{\alpha'\to 0} {\cal I}_{A_2} =
\frac{1}{X_1 X_2} + \frac{1}{(c_3+c_4-X_1)X_2}
+
\frac{1}{(c_3+c_4-X_1)(c_4+c_5-X_2)}\\
+\frac{1}{(c_3-X_1)(c_4+c_5-X_2)}
+\frac{1}{X_1(c_5-X_2)}\,.
\label{eqn:A2amp}
\end{multline}

\subsection{\texorpdfstring{\AtwoOS}{A2-OS}}

In Sec.~\ref{sec:OSdef} we found that \AtwoOSone~and \AtwoOStwo~have
the same
15 distinct cluster variables.
It is straightforward to check that they also have the
same $F$-polynomials.  Therefore, we henceforth
do not distinguish between these two cases.
The 15 $F$-polynomials of \AtwoOS~are summarized
in Tab.~\ref{tab:OSex30Fpoly}.  As always, we exclude
the trivial ``1''s that are connected to the
cluster variables in the initial cluster.  A further four
$F$-polynomials are independent of ${\bf y}$ and hence
are uninteresting (they give factors that pull out of the
stringy integral), so we set their corresponding $c_j= 0$.
\begin{table}[t]
\renewcommand{\arraystretch}{1.5}
\centering
\begin{tabular}{|c|ccccc|} \hline
    $k$ & $F_{5k+1}$ & $F_{5k+2}$ & $F_{5k+3}$ & $F_{5k+4}$ & $F_{5k+5}$ \\ \hline
    0 & 1 & 1 & $y_3 \Big[ 1 + \Big( 1 - \frac{1}{y_3} \Big) \xi_1 \xi_2 \Big]$ & $y_4 \Big[ 1 + \Big( 1 - \frac{1}{y_4} \Big) \xi_1 \xi_2 \Big]$ & $y_5$ \\
    1 & $1 - \xi_1 \xi_2$ & $1 - \xi_1 \xi_2$ & $y_3 \Big[ 1 - \frac{1}{y_3} \xi_1 \xi_2 \Big]$ & $y_4 \Big[ 1 + \Big( 2 - \frac{1}{y_4} \Big) \xi_1 \xi_2 \Big]$ & $y_5 \Big[ 1 + \xi_1 \xi_2 \Big]$ \\
    2 & $1 + \xi_1 \xi_2$ & $1 - 2 \xi_1 \xi_2$ & $y_3 \Big[ 1 - \Big( 1 + \frac{1}{y_3} \Big) \xi_1 \xi_2 \Big]$ & $y_4 \Big[ 1 - \frac{1}{y_4} \xi_1 \xi_2 \Big]$ & $y_5 \Big[ 1 + 2 \xi_1 \xi_2 \Big]$ \\ \hline
\end{tabular}
\caption{$F$-polynomials for \AtwoOS.  Here we use the notation
$y_3 = 1+y_1$, $y_4 = 1 + y_1 + y_1 y_2$ and $y_5 = 1 + y_2$; these
are the $F$-polynomials of the ordinary $A_2$ algebra;
see~(\ref{eqn:A2Fpoly}).}
\label{tab:OSex30Fpoly}
\end{table}
The super-stringy integral~\eqref{eqn:string-modified} is then
\begin{equation}
\begin{split}
    \mathcal{I}^f_{A_2^\text{OS}} = \alpha' &\int d\xi_1d\xi_2 \int_0^\infty  \frac{d y_1}{y_1} \, \frac{d y_2}{y_2} ~ y_1^{\alpha' X_1} y_2^{\alpha' X_2} y_3^{-\alpha' (c_3 + c_8 + c_{13})} y_4^{-\alpha' (c_4 + c_9 + c_{14})} y_5^{-\alpha' (c_5 + c_{10} + c_{15})} \\
    &\times \Big\{ 1 + \alpha' \Big[ ( - c_3 - c_4 - 2 c_9 - c_{10}  + c_{13} - 2 c_{15} ) \\
    &~~~~~~~~~~~~~~~ + (c_3 + c_8 + c_{13}) \frac{1}{y_3}  +(c_4 + c_9 + c_{14}) \frac{1}{y_4} \Big] \xi_1 \xi_2 \Big\}.
\end{split}
\end{equation}
Setting
\begin{equation}
    \tilde{c}_3 = c_3 + c_8 + c_{13} ~~~, \quad
    \tilde{c}_4 = c_4 + c_9 + c_{14} ~~~, \quad
    \tilde{c}_5 = c_5 + c_{10} + c_{15}
\end{equation}
we can evaluate the integrals in terms of the ordinary (bosonic)
$A_2$ stringy integral:
\begin{multline}
    \mathcal{I}^{f}_{A_2^\text{OS}} ~=~
    (-c_3-c_4-2c_9-c_{10}+c_{13}-2 c_{15})
    ~ {\cal I}_{A_2}(X_1,X_2,\tilde{c}_3, \tilde{c}_4, \tilde{c}_5) \\
    +
    \tilde{c}_3 ~ \mathcal{I}_{A_2} ( X_1, X_2, \tilde{c}_3 + \tfrac{1}{\alpha'}, \tilde{c}_4, \tilde{c}_5 ) ~+~ \tilde{c}_4 ~ \mathcal{I}_{A_2} ( X_1, X_2, \tilde{c}_3, \tilde{c}_4 + \tfrac{1}{\alpha'}, \tilde{c}_5 )\,.
\label{eqn:OSex30If}
\end{multline}
This can be evaluated explicitly in the $\alpha'\to 0$ limit
using~\eqref{eqn:A2amp}.  Note in particular that the result is finite in this limit.

Let us now analyze the polytope that describes the convergence
region for the result~\eqref{eqn:OSex30If} at nonzero $\alpha'$.
The first term converges inside the $A_2$ pentagon described in the previous
subsection, while the second and third terms
have the shifts
\begin{equation}
    \tilde{c}_3 \rightarrow \tilde{c}_3 + \tfrac{1}{\alpha'} ~~~,~~~
    \tilde{c}_4 \rightarrow \tilde{c}_4 + \tfrac{1}{\alpha'} 
\end{equation}
respectively.  Evidently for these terms, the convergence
region depends on $\alpha'$ in such a way that some edges
of the pentagon move off to infinity as $\alpha'\to 0$. The region on which all three terms converge for arbitrary $1/\alpha'$ is a rectangle: specifically, the ``bottom half'' of the pentagon drawn in Sec.~\ref{subsec:stringy-A2}.

\subsection{\texorpdfstring{$A_2^{\text{SF}}$}{A2-SF}}
\label{sec:A2SF}

Next let us take a look at the superfield version $A_2^{\text{SF}}$ of the $A_2$ cluster superalgebra. We will do the $\mathcal{N}=1$ case. For the superfield definition, there a few different possible ways to define the $F$-polynomials. We can take the coefficients in the frozen nodes
to be (A) ordinary (bosonic) variables $y_i$, or (B) superfields $Y_i=y_i(1
+ \eta \theta_i)$, as shown here:
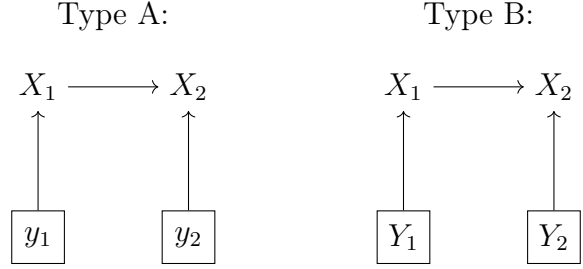
\begin{center}
\begin{tabular}{cc}
Type A: & Type B: \\[10pt]
~~~~~
\begin{tikzpicture}
\node (x1) at (0,0) {$X_1$};
\node (x2) at (2,0) {$X_2$};
\node [frozen] (y1) at (0,-2) {$y_1$};
\node [frozen] (y2) at (2,-2) {$y_2$};
\path[->]
(x1) edge (x2)
(y1) edge (x1)
(y2) edge (x2);
\end{tikzpicture} ~~~~~ & ~~~~~
\begin{tikzpicture}
\node (x1) at (0,0) {$X_1$};
\node (x2) at (2,0) {$X_2$};
\node [frozen] (y1) at (0,-2) {$Y_1$};
\node [frozen] (y2) at (2,-2) {$Y_2$};
\path[->]
(x1) edge (x2)
(y1) edge (x1)
(y2) edge (x2);
\end{tikzpicture} 
~~~~~
\end{tabular}
\end{center}
Moreover,
in the last
step of calculating the $F$-polynomials, one can choose to set (I) $X_i \rightarrow 1$ or (II) $a_i \rightarrow 1$ (recall
that $X_i = a_i(1+\eta \xi_i)$). Let us examine a few of these possibilities
case by case. 

Case (B)(I) would give the ordinary $A_2$ $F$-polynomials but with $y \rightarrow Y$, i.e. the bosonic $y$'s are promoted to superfields $Y$. That means the super-stringy integral would be
\begin{equation}
    \mathcal{I}_{A_2^\text{SF} \text{(B)(I)}} = (\alpha')^2 \int_0^\infty d y_1 \, d y_2 ~ Y_1^{\alpha' X_1 - 1} Y_2^{\alpha' X_2 - 1} Y_3^{-\alpha' c_3} Y_4^{-\alpha' c_4} Y_5^{-\alpha' c_5}\,.
\end{equation}
Substituting the $Y$-superfield expansion would give
\begin{equation}
\begin{split}
   & \mathcal{I}_{A_2^\text{SF} \text{(B)(I)}} = (\alpha')^2 \int_0^\infty  d y_1 \, d y_2 ~ y_1^{\alpha' X_1 - 1} y_2^{\alpha' X_2 - 1} y_3^{-\alpha' c_3} y_4^{-\alpha' c_4} y_5^{-\alpha' c_5} \\
    &\qquad\times \Big\{ 1 + \eta \Big[ ( \alpha' X_1 - 1 - \alpha' c_3 - \alpha' c_4 ) \theta_1 + ( \alpha' X_2 - 1 - \alpha' c_4 - \alpha' c_5 ) \theta_2 \\
    &\qquad ~~~~~~ + \alpha' c_3 \theta_1 \frac{1}{y_3} + \alpha' c_4 (\theta_1 + \theta_2) \frac{1}{y_4}   + \alpha' ( c_4 + c_5 ) \theta_2 \frac{1}{y_5} - \alpha' c_4 \theta_2 \frac{1}{y_4 y_5}  \Big] \Big\}.
\end{split}
\end{equation}
Note that
\begin{equation}
    \lim_{\alpha' \rightarrow 0} \mathcal{I}_{A_2^\text{SF} \text{(B)(I)}} = \mathcal{I}_{A_2} (X_1, X_2, c_3, c_4, c_5) \Big[ 1 - \eta (\theta_1 + \theta_2) \Big]
\end{equation}
This is the only case in which a fermionic part survives (and moreover,
is finite) in the $\alpha' \rightarrow 0$ limit. Since in this case we have $X_i \rightarrow 1$, there are no fermionic components of the $X$'s to integrate over, and we do not study an analogue of the modification~\eqref{eqn:string-modified}. This integral has the same pentagonal convergence range as the ordinary $A_2$ stringy integral.

Another interesting case appears to be
(A)(II), which would give the five $F$-polynomials
\begin{equation}
\begin{gathered}
    1 + \eta \xi_1 ~~~,~~~
    1 + \eta \xi_2 ~~~,~~~
    y_3 \Big[ 1 + \eta \Big( - \xi_1 + \frac{1}{y_3} \xi_2 \Big) \Big] ~~~, \\
    y_4 \Big[ 1 + \eta \Big( - \xi_1 - \xi_2 + \frac{1}{y_4} ( \xi_1 + \xi_2 ) \Big) \Big] ~~~,~~~ 
    y_5 \Big[ 1 + \eta \Big( - \xi_2 + ( 1 - \frac{1}{y_5} ) \xi_1 \Big) \Big]
\end{gathered}
\end{equation}
Note that since we are only setting $a_i \rightarrow 1$ instead of $X_i \rightarrow 1$, the first two $F$-polynomials are not just 1; however,
they are independent of ${\bf y}$, so we do not need to
include them in the super-stringy integral (they pull out as
overall, uninteresting factors). Therefore we consider
\begin{equation}
\begin{split}
    \mathcal{I}_{A_2^\text{SF} \text{(A)(II)}} = (\alpha')^2 \int_0^\infty & \frac{d y_1}{y_1} \, \frac{d y_2}{y_2} ~ y_1^{\alpha' X_1} y_2^{\alpha' X_2 } y_3^{-\alpha' c_3} y_4^{-\alpha' c_4} y_5^{-\alpha' c_5} \\
    &\times \Big\{ 1 - \alpha' \eta \Big[ ( - c_3 - c_4 + c_5 ) \xi_1 + ( - c_4 - c_5 ) \xi_2\label{eq:z} \\
    &~~~~~~~~~~~~~~~ + c_3 \xi_2 \frac{1}{y_3} + c_4 (\xi_1 + \xi_2) \frac{1}{y_4} - c_5 \xi_1 \frac{1}{y_5} \Big] \Big\}.
\end{split}
\end{equation}
Since all the fermionic terms vanish as $\alpha' \rightarrow 0$ in this case, let us study a modification already advertised just below the integral~\eqref{eqn:string-modified}.  Replacing
$(\alpha')^2$ in~\eqref{eq:z} by $\alpha' \int d\xi_1$ and
evaluating the $y$-integrals using the definition~\eqref{eq:pentagonintegral} gives
\begin{multline}
    \mathcal{I}^{f (1)}_{A_2^\text{SF} \text{(A)(II)}} =~ \eta \Big[
    (c_3+c_4-c_5) \mathcal{I}_{A_2} (X_1,X_2,c_3,c_4,c_5)
    \\
    - c_4 \, \mathcal{I}_{A_2} (X_1, X_2, c_3, c_4 + \tfrac{1}{\alpha'}, c_5) + c_5 \, \mathcal{I}_{A_2} (X_1, X_2, c_3, c_4, c_5 + \tfrac{1}{\alpha'}) \Big]. \label{eqn:SFexAII1If} 
    \end{multline}
On the other hand, replacing $(\alpha')^2$
by $\alpha' \int d\xi_2$ leads to
    \begin{multline}
    \mathcal{I}^{f (2)}_{A_2^\text{SF} \text{(A)(II)}} =~ \eta \Big[
    (c_4+c_5) \mathcal{I}_{A_2}(X_1,X_2,c_3,c_4,c_5)
    \\
    - c_3 \, \mathcal{I}_{A_2} (X_1, X_2, c_3 + \tfrac{1}{\alpha'}, c_4, c_5) - c_4 \, \mathcal{I}_{A_2} (X_1, X_2, c_3, c_4 + \tfrac{1}{\alpha'}, c_5) \Big].  \label{eqn:SFexAII2If}
\end{multline}
These results are qualitatively similar to those in the
previous subsection: in each case we get a sum of three terms,
in two of which the convergence region is changed by shifting some
edge(s) of the pentagon outward by $+1/\alpha'$, and each of the results is finite in the limit $\alpha'\to 0$.

\acknowledgments

We are grateful to Nima Arkani-Hamed, Yu-tin Huang, Lecheng Ren, Martin Ro\v{c}ek and Giulio Salvatori for helpful correspondence, and to Anders Schreiber for collaboration in the initial stage of this work.  This work was supported in part by the US Department of Energy under contract {DE}-{SC}0010010 Task A (MS, AV) and by Simons Investigator Award \#376208 (AV).
The research of S. J. Gates, Jr. and S.-N. H. Mak is supported by the endowment from the Ford Foundation Professorship of Physics at Brown University and they gratefully acknowledge the support of the Brown Theoretical Physics Center. Part of S.-N. H. Mak's work is also supported by the Galkin Foundation Fellowship at Brown University.

\appendix

\section{More examples: superfield definition}
\label{appen:SFex-other}

\subsection{\texorpdfstring{$B_2^{\text{SF}}$ and $C_2^{\text{SF}}$}{B2-SF and C2-SF}}

The $B_2$ cluster variables are
\begin{equation}
\begin{gathered}
    X_{1} ~~~,~~~
    X_{2} ~~~,~~~
    X_{3} = \frac{1 + X_{2}}{X_{1}} ~~~,~~~
    X_{4} = \frac{X_{1}^{2} + (1 + X_{2})^{2}}{X_{1}^{2} X_{2}} ~~~, \\
    X_{5} = \frac{1 + X_{1}^{2} + X_{2}}{X_{1} X_{2}} ~~~,~~~
    X_{6} = \frac{1 + X_{1}^{2}}{X_{2}} \,.
\end{gathered}
\end{equation}
If we start with $\mathcal{N}=1$ initial cluster variables,
and define $X_i = a_i ( 1 + \eta \, \theta_i)$, 
the fermionic components will be

\begin{equation}
\begin{split}
    \theta_3 ~=&~ -~ \theta_1 ~+~ \frac{a_2}{1 + a_2} \, \theta_2  \\
    \theta_4 ~=&~ -~ \frac{2 (1 + a_2)^2}{a_1^2 + (1 + a_2)^2} \, \theta_1 ~-~ \frac{1 + a_1^2 - a_2^2}{a_1^2 + (1 + a_2)^2} \, \theta_2  \\
    \theta_5 ~=&~ -~ \frac{1 - a_1^2 + a_2}{1 + a_1^2 + a_2} \, \theta_1 ~-~ \frac{1 + a_1^2}{1 + a_1^2 + a_2} \, \theta_2  \\
    \theta_6 ~=&~ \frac{2 a_1^2}{1 + a_1^2} \, \theta_1 ~-~ \theta_2\,.
\end{split}
\end{equation}
Note that $C_2$ cluster variables can be obtained from those of $B_2$ by exchanging $X_1 \leftrightarrow X_2$ (and reversing the order of the 6 variables).

\subsection{\texorpdfstring{$A_3^{\text{SF}}$}{A3-SF}}

The $A_3$ cluster variables are
\begin{equation}
\begin{gathered}
    X_{1} ~~~,~~~
    X_{2} ~~~,~~~
    X_{3} ~~~,~~~
    X_{4} = \frac{1 + X_{2}}{X_{1}} ~~~,~~~
    X_{5} = \frac{X_{1} + X_{3}}{X_{2}} ~~~,~~~
    X_{6} = \frac{1 + X_{2}}{X_{3}} ~~~, \\
    X_{7} = \frac{X_{1} + (1 + X_{2}) X_{3}}{X_{1} X_{2}} ~~~,~~~
    X_{8} = \frac{X_{3} + (1 + X_{2}) X_{1}}{X_{2} X_{3}} ~~~,~~~
    X_{9} = \frac{(1 + X_{2})(X_{1} + X_{3})}{X_{1} X_{2} X_{3}} 
\end{gathered}
\end{equation}
If we start with $\mathcal{N}=1$ initial cluster variables
and define $X_{i} = a_{i} ( 1 + \eta \, \theta_{i} )$,
the fermionic components will be 

\begin{equation}
\begin{split}
    \theta_4 ~=&~ -~ \theta_1 ~+~ \frac{a_2}{1 + a_2} \, \theta_2 \\
    \theta_5 ~=&~ \frac{a_1}{a_1 + a_3} \, \theta_1 ~-~ \theta_2 ~+~ \frac{a_3}{a_1 + a_3} \, \theta_3 \\
    \theta_6 ~=&~ \frac{a_2}{1 + a_2} \, \theta_2 ~-~ \theta_3 \\
    \theta_7 ~=&~ -~ \frac{(1 + a_2) a_3}{a_1 + (1 + a_2) a_3} \, \theta_1 ~-~ \frac{a_1 + a_3}{a_1 + (1 + a_2) a_3} \, \theta_2 ~+~ \frac{(1 + a_2) a_3}{a_1 + (1 + a_2) a_3} \, \theta_3 \\
    \theta_8 ~=&~ \frac{a_1 (1 + a_2)}{a_1 (1 + a_2) + a_3} \, \theta_1 ~-~ \frac{a_1 + a_3}{a_1 (1 + a_2) + a_3} \, \theta_2 ~-~ \frac{a_1 (1 + a_2)}{a_1 (1 + a_2) + a_3} \, \theta_3 \\
    \theta_9 ~=&~ -~ \frac{a_3}{a_1 + a_3} \, \theta_1 ~-~ \frac{1}{1 + a_2} \, \theta_2 ~-~ \frac{a_1}{a_1 + a_3} \, \theta_3 \,.
\end{split}
\end{equation}


\begin{thebibliography}{99}

\bibitem{FZ1}
S.~Fomin and A.~Zelevinsky,
``Cluster algebras I: Foundations'',
Journal of the American Mathematical Society {\bfseries 15}, 497 (2002)
[arXiv:math/0104151].

\bibitem{clusterbk}
S.~Fomin, L.~Williams and A.~Zelevinsky,
``Introduction to Cluster Algebras,''
[arXiv:1608.05735 [math.CO]], [arXiv:1707.07190 [math.CO]], [arXiv:2008.09189 [math.CO]].

\bibitem{OS2015}
V.~Ovsienko,
``Cluster superalgebras,''
[arXiv:1503.01894 [math.CO]].

\bibitem{LMRS2017}
L.~Li, J.~Mixco, B.~Ransingh and A.~K.~Srivastava,
``An Introduction to Supersymmetric Cluster Algebras,''
[arXiv:1708.03851 [math.RA]].

\bibitem{OS2018}
V.~Ovsienko and M.~Shapiro,
``Cluster algebras with Grassmann variables,''
[arXiv:1809.01860 [math.CO]].

\bibitem{SV2019}
E.~Shemyakova and T.~Voronov,
``On super Pl\"ucker embedding and cluster algebras,''
[arXiv:1906.12011 [math.DG]].

\bibitem{Golden:2013xva}
J.~Golden, A.~B.~Goncharov, M.~Spradlin, C.~Vergu and A.~Volovich,
``Motivic Amplitudes and Cluster Coordinates,''
JHEP \textbf{01}, 091 (2014)
[arXiv:1305.1617 [hep-th]].

\bibitem{Arkani-Hamed:2012zlh}
N.~Arkani-Hamed, J.~L.~Bourjaily, F.~Cachazo, A.~B.~Goncharov, A.~Postnikov and J.~Trnka,
``Grassmannian Geometry of Scattering Amplitudes,''
[arXiv:1212.5605 [hep-th]].

\bibitem{Arkani-Hamed:2019mrd}
N.~Arkani-Hamed, S.~He and T.~Lam,
``Stringy canonical forms,''
JHEP \textbf{02}, 069 (2021)
[arXiv:1912.08707 [hep-th]].

\bibitem{Koba:1969kh}
Z.~Koba and H.~B.~Nielsen,
``Manifestly crossing invariant parametrization of $n$ meson amplitude,''
Nucl.\ Phys.\ B {\bf 12}, 517 (1969).

\bibitem{Arkani-Hamed:2019vag}
N.~Arkani-Hamed, S.~He, G.~Salvatori and H.~Thomas,
``Causal Diamonds, Cluster Polytopes and Scattering Amplitudes,''
[arXiv:1912.12948 [hep-th]].

\bibitem{Chicherin:2020umh}
D.~Chicherin, J.~M.~Henn and G.~Papathanasiou,
``Cluster algebras for Feynman integrals,''
Phys. Rev. Lett. \textbf{126}, no.9, 091603 (2021)
[arXiv:2012.12285 [hep-th]].

\bibitem{Banerjee:2018tun}
P.~Banerjee, A.~Laddha and P.~Raman,
``Stokes polytopes: the positive geometry for $\phi^{4}$ interactions,''
JHEP \textbf{08}, 067 (2019)
[arXiv:1811.05904 [hep-th]].

\bibitem{Salvatori:2018fjp}
G.~Salvatori and S.~L.~Cacciatori,
``Hyperbolic Geometry and Amplituhedra in 1+2 dimensions,''
JHEP \textbf{08}, 167 (2018)
[arXiv:1803.05809 [hep-th]].

\bibitem{Salvatori:2018aha}
G.~Salvatori,
``1-loop Amplitudes from the Halohedron,''
JHEP \textbf{12}, 074 (2019)
[arXiv:1806.01842 [hep-th]].

\bibitem{MOZ2021a}
G.~Musiker, N.~Ovenhouse and S.~W.~Zhang,
``An Expansion Formula for Decorated Super-Teichm\"uller Spaces,''
SIGMA \textbf{17}, 080 (2021)
[arXiv:2102.09143 [math.CO]].

\bibitem{MOZ2021b}
G.~Musiker, N.~Ovenhouse and S.~W.~Zhang,
``Double Dimer Covers on Snake Graphs from Super Cluster Expansions,''
[arXiv:2110.06497 [math.CO]].

\bibitem{Abel}
N.~H.~Abel,
``Note sur la fonction $\psi(x) = x + \frac{x^2}{2^2} + \frac{x^3}{3^2} + \cdots + \frac{x^n}{n^2} + \cdots$,'' in
\emph{{\OE}uvres compl\`etes de Niels Henrik Abel, Tome II},
pp.~189-193, Oslo, 1881.

\bibitem{Gauss}
C.~F.~Gauss,
``Pentagramma mirificum,'' in
\emph{Werke, Band III}, pp.~481--490, G\"ottingen, 1863.

\bibitem{Gangl} H.~Gangl,
``The Grassmannian complex and Goncharov's motivic complex in weight 4,''
arXiv:1801.07816 [math.NT].

\bibitem{CGR}
S.~Charlton, H.~Gangl and D.~Radchenko,
``Functional equations of polygonal type for mulitiple polylogarithms in weights 5, 6 and 7,''
arXiv:2012.09840 [math.NT].

\bibitem{GR}
A.~B.~Goncharov and D.~Rudenko,
``Motivic correlators, cluster varities and Zagier's conjecture on $\zeta_F(4)$,''
arXiv:1803.08585.

\bibitem{Arkani-Hamed:2017tmz}
N.~Arkani-Hamed, Y.~Bai and T.~Lam,
``Positive Geometries and Canonical Forms,''
JHEP \textbf{11}, 039 (2017)
[arXiv:1703.04541 [hep-th]].

\bibitem{FZ4}
S.~Fomin and A.~Zelevinsky,
``Cluster algebras IV: Coefficients'',
Compositio Mathematica {\bf series 143} no.~1, 112 (2007),
[arXiv:math/0602259].

\bibitem{Arkani-Hamed:2019plo}
N.~Arkani-Hamed, S.~He, T.~Lam and H.~Thomas,
``Binary Geometries, Generalized Particles and Strings, and Cluster Algebras,''
[arXiv:1912.11764 [hep-th]].

\bibitem{He:2020ray}
S.~He, L.~Ren and Y.~Zhang,
``Notes on polytopes, amplitudes and boundary configurations for Grassmannian string integrals,''
JHEP \textbf{04}, 140 (2020)
[arXiv:2001.09603 [hep-th]].

\bibitem{He:2020onr}
S.~He, Z.~Li, P.~Raman and C.~Zhang,
``Stringy canonical forms and binary geometries from associahedra, cyclohedra and generalized permutohedra,''
JHEP \textbf{10}, 054 (2020)
[arXiv:2005.07395 [hep-th]].

\bibitem{Arkani-Hamed:2020tuz}
N.~Arkani-Hamed, S.~He and T.~Lam,
``Cluster Configuration Spaces of Finite Type,''
SIGMA \textbf{17}, 092 (2021)
[arXiv:2005.11419 [math.AG]].

\bibitem{He:2021zuv}
S.~He, Y.~Wang, Y.~Zhang and P.~Zhao,
``Notes on worldsheet-like variables for cluster configuration spaces,''
[arXiv:2109.13900 [hep-th]].

\end{thebibliography}
\end{document}